\documentclass[conference]{IEEEtran}
\usepackage[bookmarks=false]{hyperref}
\usepackage{cite}
\usepackage{amsmath,amssymb,amsfonts}
\usepackage{algorithmic}
\usepackage{graphicx}
\usepackage{orcidlink}
\usepackage{textcomp}
\usepackage{xcolor}
\usepackage{siunitx}
\usepackage{multirow}

\def\BibTeX{{\rm B\kern-.05em{\sc i\kern-.025em b}\kern-.08em
    T\kern-.1667em\lower.7ex\hbox{E}\kern-.125emX}}

\makeatletter
\def\ps@IEEEtitlepagestyle{%
  \def\@oddfoot{\hbox{}%
    \scriptsize
    \hfil
    This work has been submitted to the IEEE for possible publication. 
    Copyright may be transferred without notice, after which this version 
    may no longer be accessible.\hfil}%
  \def\@evenfoot{}%
}
\makeatother

\begin{document}

\title{BeamSeek: Deep Learning-based DOA Estimation for Low-Complexity mmWave Phased Arrays}

\author{
    \IEEEauthorblockN{
        Arav Sharma$^1$ \orcidlink{0009-0004-7928-7047},
        Lei Chi$^1$ \orcidlink{0009-0002-1462-4319},
        Ari Gebhardt$^1$ \orcidlink{0009-0006-4991-5617},
        Alon S. Levin$^2$ \orcidlink{0000-0001-7579-9214},
        Timothy R. Hoerning$^1$
        and Sam Keene$^1$
    }
    \IEEEauthorblockA{
        $^1$The Cooper Union for the Advancement of Science and Art, New York City, USA \\
        $^2$Department of Electrical Engineering, Columbia University, New York City, USA \\
        \{arav.sharma, lei.chi, ari.gebhardt, timothy.hoerning, keene\}@cooper.edu, alon.s.levin@columbia.edu
    }
}

\maketitle

\begin{abstract}
A novel approach combining agile beam switching with deep learning to enhance the speed and accuracy of Direction of Arrival (DOA) estimation for millimeter-wave (mmWave) phased array systems with low-complexity hardware implementations is proposed and evaluated. Traditional DOA methods requiring direct access to individual antenna elements are impractical for analog or hybrid beamforming systems prevalent in modern mmWave implementations. Recent agile beam switching techniques have demonstrated rapid DOA estimation, but their accuracy and robustness can be further improved via deep learning. BeamSeek addresses these limitations by employing a Multi-Layer Perceptron (MLP) and specialized data augmentation that emulates real-world propagation conditions. The proposed approach was experimentally validated at 60 GHz using the NSF PAWR COSMOS testbed, demonstrating significant improvements over a correlation-based method across various Signal-to-Noise Ratio (SNR) levels. Results show that BeamSeek achieves up to an $\bf{8^\circ}$ reduction in average estimation error compared to this baseline, with particular advantages in noisy channels. This makes it especially suitable for practical mmWave deployments in environments characterized by multipath interference and hardware constraints.
\end{abstract}


\section{Introduction}
Millimeter-wave (mmWave) technology is a crucial component of next-generation wireless communications systems, offering greater bandwidth and data capacity to meet growing demands \cite{rappaport2013millimeter}. A significant challenge in mmWave implementations is their susceptibility to substantial path loss and signal blockage, which demands highly directional beamforming to ensure reliable communication links \cite{Wang2018}. Enabling this requires accurate Direction of Arrival (DOA) estimation to optimize beam steering.

\begin{figure}
    \centering
    \includegraphics[width=0.8\linewidth]{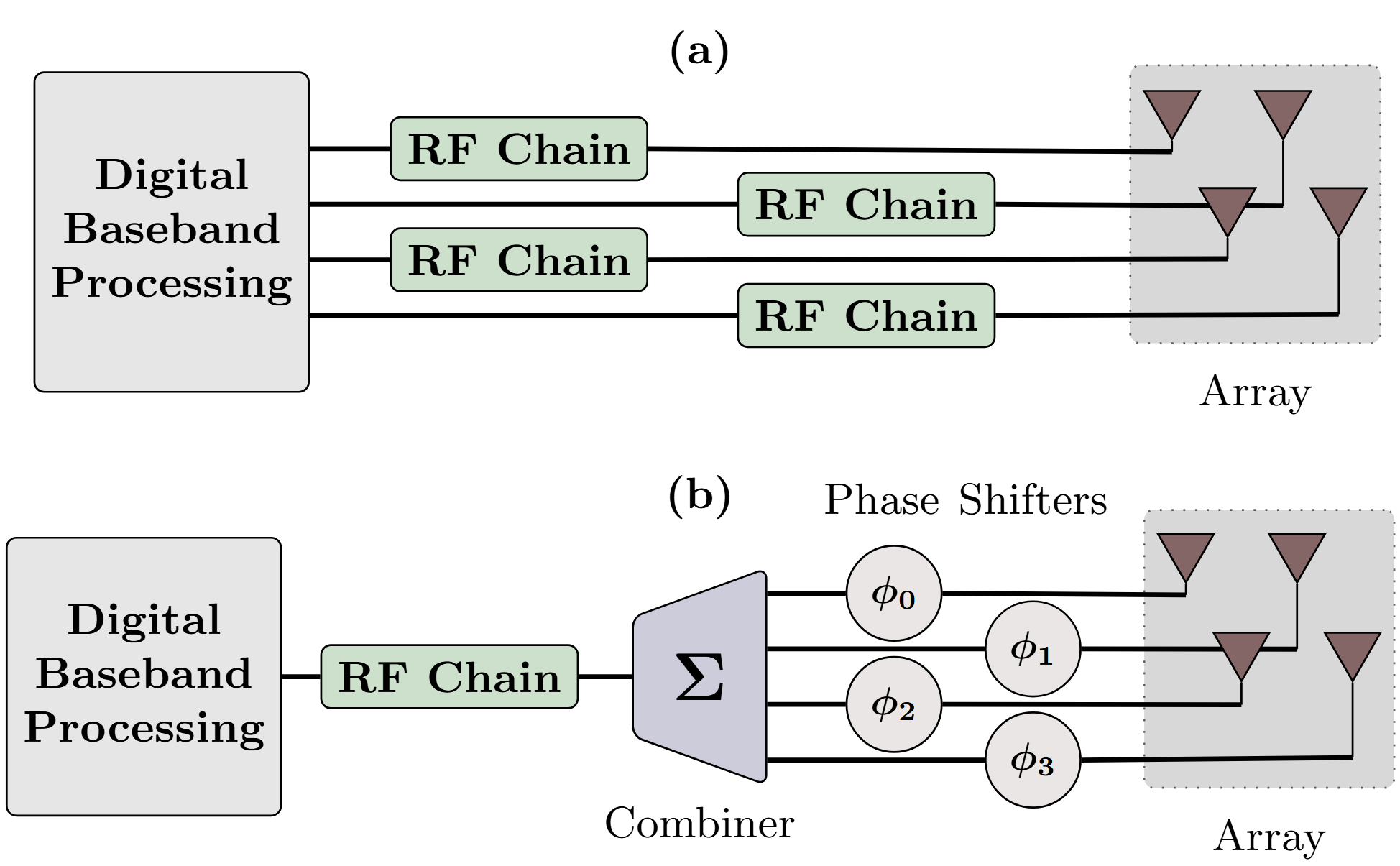}
    \caption{Digital beamforming with multiple RF chains (a) vs. analog beamforming with a single RF chain (b).}
    \label{fig:analogVsBeam}
\end{figure}

Traditional DOA estimation approaches for phased array systems rely on methods such as MUSIC (MUltiple SIgnal Classification) \cite{schmidt1986multiple} and ESPRIT (Estimation of Signal Parameters via Rotational Invariance Techniques) \cite{roy1989esprit}. These techniques require direct access to individual antenna element data, which is only available in arrays with dedicated RF chains for each element. However, for many practical mmWave systems, analog beamforming architectures with a single RF chain are often preferred due to their reduced hardware complexity, power consumption, and cost \cite{Choi2016}. For example, a majority of 5G gNodeB installments currently rely on a limited number of RF chains \cite{Gallyas-Sanhueza2024}. The difference in these beamforming architectures is shown in Figure~\ref{fig:analogVsBeam}.

\begin{figure*}[h] \centering \includegraphics[width=0.8\linewidth]{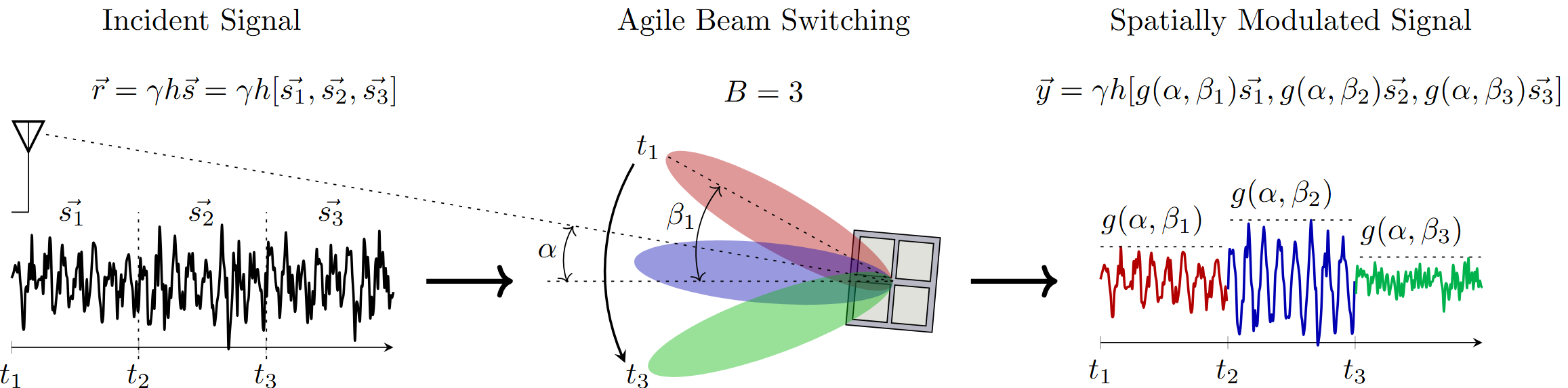} \caption{System model used to derive the input data, demonstrating spatial modulation of the incident signal through agile beam switching for $B=3$.} \label{fig:sysdia} \end{figure*}

Under these hardware constraints, only the combined output from the antenna array is accessible, presenting significant challenges for conventional DOA estimation methods. The simplest approach is beam sweeping, where the array sequentially scans all angular directions and chooses the beam corresponding to the maximum received power\cite{giordani2018tutorial}. While straightforward to implement, this approach introduces latency proportional to the angular resolution, limiting its applicability in dynamic environments.

An innovative solution to this challenge was introduced by Gallyas-Sanhueza et al. \cite{Gallyas-Sanhueza2024}, who developed an agile beam switching technique capable of rapidly estimating DOA within a single OFDM symbol. While effective in ideal conditions, this approach may face challenges in high noise and multipath scenarios. Additionally, this method relies on a known pilot signal, which potentially limits its applications. The primary contributions of this work include:
\begin{itemize}
    \item A novel approach that combines agile beam switching with Deep Learning (DL) to enhance DOA estimation accuracy for hardware-constrained phased arrays, experimentally validated at 60 GHz on the National Science Foundation (NSF) Platforms for Advanced Wireless Research (PAWR) Cloud Enhanced Open Software Defined Mobile Wireless Testbed for City-Scale Deployment.
    \item A data augmentation method informed by real-world propagation behavior to increase the robustness of BeamSeek and expand the limited training set.
    \item A pilot-free correlation-based DOA estimation method to use a baseline for comparison.
\end{itemize}

\section{Related Work}

\subsection{Hierarchical Beam Scanning}

Hierarchical beam scanning techniques effectively mitigate the latency associated with exhaustive beam scanning by employing progressively narrower beams to systematically reduce the search space \cite{xiao2016hierarchical}. However, these approaches rely on a multi-stage search process, which introduces latency of its own. 
Additionally, these techniques suffer in rapidly changing environments, such as high-mobility scenarios or settings with dynamic obstacles, due to the inherent delay in iterative refinement of the beam code book.

\subsection{Agile Beam Switching}
Recent advances in phased array technology enable beam switching within microseconds. Gallyas-Sanhueza et al. \cite{Gallyas-Sanhueza2024} utilized agile switching to  modulate a received pilot signal with a spatially varying pattern and correlated it with theoretical patterns to estimate the DOA. This approach reduces latency and requires only one RF chain, though its real-world performance remains understudied. An implementation derived from this method that does not rely on a known pilot serves as a baseline to evaluate the performance of the approach introduced in this work.

\subsection{Deep Learning for DOA Estimation}
Deep learning (DL) approaches have shown promising results in addressing the limitations of traditional DOA estimation methods. Hassan et al. \cite{hassan2024deep} showed that convolutional neural networks (CNN) for DOA estimation outperform MUSIC in low SNR and multipath scenarios. Similarly, Huang et al. \cite{huang2020deep} proposed a deep neural network for DOA estimation with improved robustness to calibration errors. However, most DL approaches have focused on processing data from digital architectures with access to individual antenna elements. The application of DL techniques to hardware-constrained analog beamforming systems with a limited number of RF chains presents a significant research opportunity. In addition, many proposed DL methods have only been tested in simulation. Experimentally verifying the DL method in this work on a real-world testbed provides additional context regarding its deployability.

\section{Methodology}

\subsection{Input Data}

We consider a signal incident on a uniform planar array (UPA), originating from a transmitter located at azimuthal angle $\alpha$. The colatitude of the transmitter ($\phi$) is not considered due to restrictions of the testbed. Let ${\vec{s}} = [s_1, s_2, \ldots, s_N]$ represent the transmitted signal of length $N$, where each element $s_n \in \mathbb{C}$ represents a sampled data point. The magnitude of the transmitted signal is scaled by a constant $\gamma \in \mathbb{R}$, determined by the power of the transmitter. To model the incident signal, we apply a channel coefficient $h \in \mathbb{C}$, which represents the phase shift and attenuation due to propagation through the medium.  The incident signal, $\vec{r}$ can then be expressed as $\vec{r} = \gamma h \vec{s} = \gamma h [s_1, s_2, ..., s_N]$. While receiving the signal, we rapidly switch the steering angle of the UPA to $B$ equally spaced beams. Due to the short period of observation, we can assume that $\gamma$ and $h$ remain constant. The incident signal is segmented and rewritten as
\begin{equation}
    \vec{r} = \gamma h [\vec{s}_1, \vec{s}_2, ..., \vec{s}_B]\text{,}
\end{equation}
\noindent where each component $\vec{s_n} \in \mathbb{C}^{N/B}$ is an $N/B$ length segment of the transmitted signal and is received by a corresponding beam with a steering angle $\beta_n$. As a consequence of beamforming, the gain of each $\vec{s_n}$  varies strongly based on values of $\alpha$ and $\beta_n$, and can denoted by a coefficient $g(\alpha, \beta_n) \in  \mathbb{C}$. By rapidly switching $\beta$, we modulate $\vec{r}$ by a spatially dependent pattern, as show in Figure~\ref{fig:sysdia}. This modulated signal can be expressed as 
\begin{equation}
    \vec{y} = \gamma h [g(\alpha, \beta_1)\vec{s_1}, g(\alpha, \beta_2)\vec{s_2}, \ldots, g(\alpha, \beta_B)\vec{s_B}]\text{,}
\end{equation} which can be written as
$\vec{y} = [\vec{y_1}, \vec{y_2}, \ldots, \vec{y_B}]$,
where each $\vec{y_n} = \gamma h g(\alpha, \beta_n)\vec{s_n}$.
The power of $\vec{y_n}$ can be calculated as 
\begin{equation}
P(\vec{y}_n)=\frac{1}{N/B}\left\lVert\vec{y}_n\right\rVert^2\text{,}
\end{equation} which is equivalent to
\begin{equation}
      P(\vec{y}_n) = |\gamma hg(\alpha, \beta_n)|^2P(\vec{s}_n) \text{.}
\end{equation}
It is important to note that $\gamma$, $h$, and $P(\vec{s}_n)$ are constant during the period of observation. Therefore, $P(\vec{y}_n)$ is directly proportional to $|g(\alpha, \beta_n)|^2$, which removes the dependence of our DOA estimation methods on phase and by extension a known pilot. Finally, we can define the power profile vector
\begin{equation}
\vec{P_\alpha} = [P(\vec{y}_1), P(\vec{y}_2), \ldots, P(\vec{y}_B)] \text{,}
\end{equation}
which will serve as the input to our DOA estimation methods.

\subsection{Correlation-Based DOA Estimation}
The correlation-based method generates theoretical power profiles for each candidate DOA $ \alpha'$ and selects the estimated DOA $\hat{\alpha}$ that exhibits the highest correlation with the input power profile vector. To calculate our theoretical power profiles, we must calculate $\vec{P}_{\alpha'}= [|g(\alpha', \beta_1)|^2, |g(\alpha', \beta_2)|^2, \ldots, |g(\alpha', \beta_B)|^2]$ for each $\alpha'$ and $\beta_n$. The candidate angles are constrained to the range of $\pm45^\circ$, due to the scan range of the device. For a given signal $\vec s$ and steering vectors $\vec{\alpha}'$ and $\vec{\beta}_n$, we can model the signal incident on the UPA as outer-product ${\vec s}^T\vec\alpha'$ and the signal received by steering angle $\beta_n$ as ${\vec s}^T\vec\alpha'\vec\beta^H$. This results in the signal being scaled by the inner-product of $\vec\alpha'$ and $\vec\beta_n$, yielding the derivation
\begin{equation}
    g(\alpha', \beta_n) = \vec\alpha'\vec\beta_n^H \text{,}
\end{equation}
which can be used to generate the theoretical power profiles. To select $\hat{\alpha}$ we maximize the cosine similarity between both power profile vectors: 
\begin{equation}
    \hat{\alpha} = \operatorname*{arg\,max}_{\alpha' \in [-45^\circ, \,45^\circ]} \frac{\vec{P_{\alpha}} \cdot \vec{P_{\alpha'}^T}}{\lVert \vec{P_\alpha} \rVert \lVert \vec{P_{\alpha'}}\rVert}\text{.}
\end{equation}
Cosine similarity was chosen as the metric because it does not consider the magnitude of the vectors, which is significant because $\vec{P_{\alpha}}$ is scaled by unknown constant $|\gamma hg(\alpha, \beta_n)|^2P(\mathbf{s}_n)$.

\subsection{BeamSeek}

\begin{figure}
    \centering
    \includegraphics[width=0.85\linewidth]{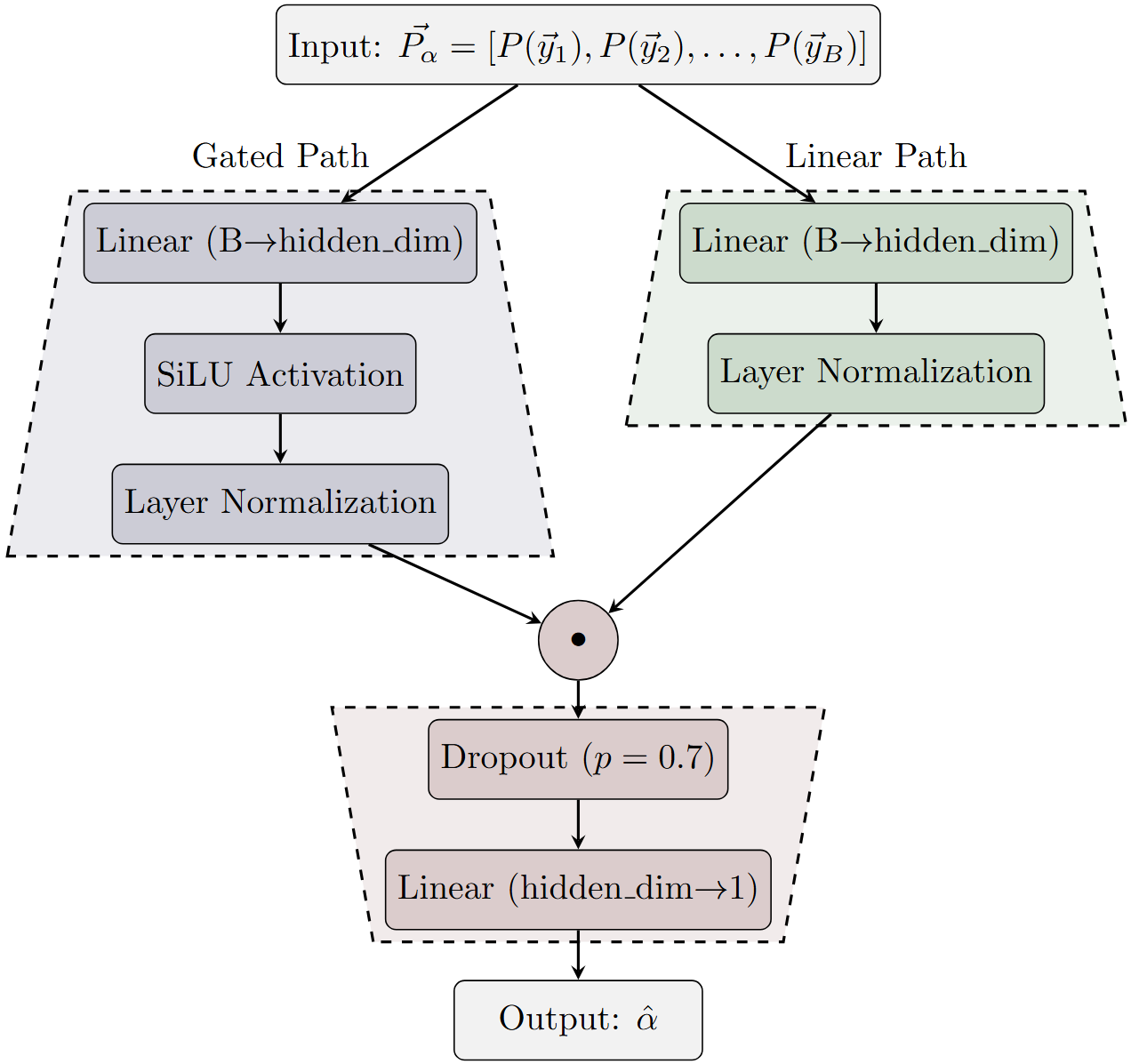}
    \caption{Architecture of the BeamSeek SwiGLU neural network for DOA estimation, featuring parallel gated and linear paths that process beam gain profiles through SiLU activation and layer normalization before element-wise multiplication, dropout regularization, and final angle prediction.}
    \label{fig:beamseek}
\end{figure}

\subsubsection{Neural Network Architecture}

Our approach employs a Multi-Layer Perceptron (MLP) as shown in Figure \ref{fig:beamseek} to predict angles from power profiles. The network receives power profiles of dimension B as input and outputs a single angle prediction. We use mean squared error as our loss function, which penalizes larger errors more severely than smaller ones, encouraging high precision in angle predictions.

The MLP architecture consists of a single SwiGLU gated linear unit (GLU) feed-forward network (FFN) with a hidden dimension of 384. SwiGLU, or Swish-Gated Linear Unit, combines the Swish activation function with a gating mechanism similar to GLUs \cite{Shazeer2020}. Mathematically, the SwiGLU is defined as
\begin{equation}
\text{FFN}_\text{SwiGLU}(x, W, V, W_2) = (\text{Swish}_1(xW) \otimes xV)W_2 \text{,}
\end{equation}
where $x$ is the input vector, $W$, $V$, and $W_2$ are weight matrices, $\otimes$ represents element-wise multiplication, and the Swish activation function is defined as
\begin{equation}
\text{Swish}_1(x) =\text{SiLU}(x) = x \cdot \text{sigmoid}(x)\text{.}
\end{equation}
This activation approach enables more expressive feature transformations by enabling the network to modulate information flow through the multiplication of two separate branches of computation. One branch applies the Swish activation while the other learns a gating function that controls which information passes through. This structure enables more complex functional mappings while maintaining efficient gradient flow during training.

\subsubsection{Data Augmentation and Iterative Training}
\begin{figure*}
    \centering
    \includegraphics[width=0.85\linewidth]{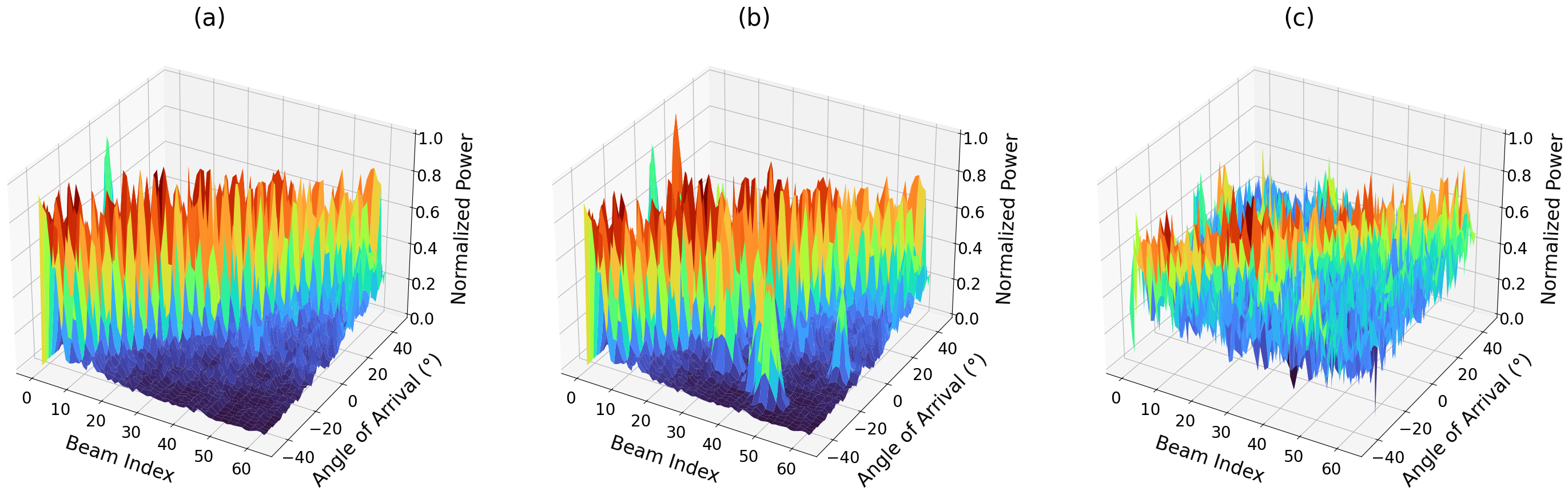}
    \caption{The augmentation process of base training data including the LOS power profile $\vec{P}_{LOS}$ (a), the data after the addition of spatially correlated multipath components $\vec{P}_{TOT}$ (b), and the final training data after the addition of white Gaussian noise (c) used for one iteration of training.}
    \label{fig:augmentation}
\end{figure*}

\begin{figure}[b!]
    \centering
    \includegraphics[width=0.85\linewidth]{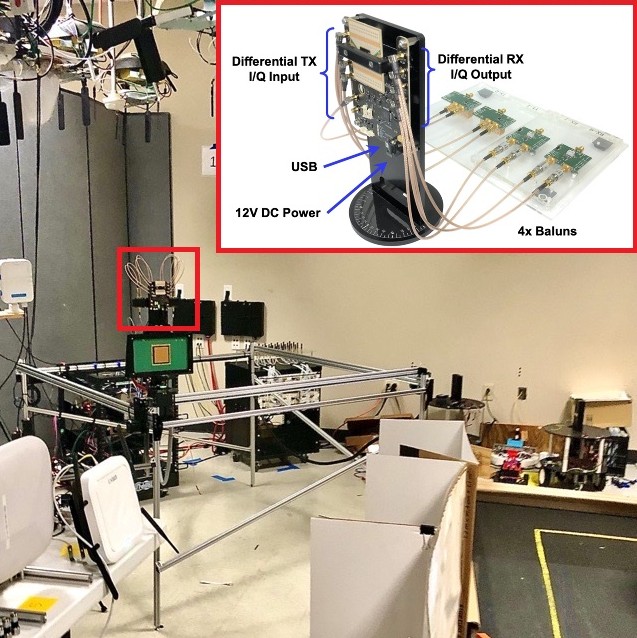}
    \caption{Sivers IMA transceiver mounted on moving XY table.}
    \label{fig:table}
\end{figure}

To prevent overfitting and improve generalization, we implement several regularization techniques, including layer normalization, L2 weight regularization with a coefficient of $10^{-5}$, and a dropout rate of 0.7 between layers. For training, we process data in batches of 256 examples using the Adam optimizer \cite{kingma2014adam}, which adapts learning rates for each parameter based on gradient estimates. A learning rate scheduler gradually reduces the rate from 0.1 to 0.01 over 3,750 iterations of unique augmented versions of the dataset.

BeamSeek relies on an exhaustive calibration scan of all beams and all angles, differentiating it from the baseline correlation-based approach. To compensate for the limited data in a single scan, we use data augmentation techniques to expand the scan into a much larger training set. Our data augmentations emulate perturbations in real-world environments, including additive white noise and multipath interference, training the neural network to be more robust and accurate. Every training iteration, new augmented data is passed to the FFN.
The foundational data for the iterative training method was a high-SNR (39.9 dB) beam scan, allowing a more accurate estimate of the SNR after augmentation, as well as forming a strong base for augmentation and training due to a greater density of features pertinent to the phased array's response to a given incident signal. \par
MmWave phased arrays are specifically designed for Line-Of-Sight (LOS) wireless communications, so the multipath effects are modeled as Rician. Normally, the Non-Line-of-Sight (NLOS) components are independently sampled from a Rayleigh distribution as it has been shown that multipath contributions, particularly in an urban environment, follow this distribution \cite{Rappaport2002}. In this case, independently adding noise from a Rayleigh distribution across the power profiles  would not be suitable because the distortion caused by NLOS signals should be spatially correlated \cite{Durgin2000}; a strong peak outside of LOS should also be seen by the adjacent beams. The phase shift caused by multipath interference does not need to be considered because BeamSeek only processes power profiles, which are unaffected by phase. Therefore, the sum of Gaussian spatial distributions is an appropriate model to use here. The scattered NLOS peaks are generated as

\begin{equation}
    \vec{P}_{NLOS} = \sum_{i=1}^{M} P_i \exp\bigg(-\frac{(\alpha - \alpha_i)^2}{2\sigma_\alpha^2} -\frac{(\beta - \beta_i)^2}{2\sigma_\beta^2}\bigg) \text{,}
\end{equation}

\noindent where $\vec{P}_{NLOS}$ is the power profile of the NLOS components as a function of $\alpha$ and $\beta$, $M$ is the number of NLOS components, $P_i$ is the power contribution of the \emph{i}th NLOS component, $\alpha_i$ and $\beta_i$ are the angle and beam index of the NLOS component, and $\sigma_{\alpha}$ and $\sigma_{\beta}$ are the spread parameters that shape the the NLOS peak. These parameters are randomly chosen during augmentation. For a large $M$, $\vec{P}_{NLOS}$ will follow a joint Gaussian distribution because it is the sum of many small Gaussian-distributed components. The envelope of a joint Gaussian distribution is a Rayleigh distribution, so our augmentation model can approximate conventional Rayleigh fading models in cases with a large $M$. \par The generated NLOS power profile is combined with the LOS power profile according to the Rician model:

\begin{equation}
    \vec{P}_{TOT} = \frac{K}{K+1}\vec{P}_{LOS} + \frac{1}{K+1}\vec{P}_{NLOS} \text{,}
\end{equation}

\noindent where $\vec{P}_{TOT}$ is the total received power, $K$ is the Rician factor, $\vec{P}_{LOS}$ is the LOS power profile, and $\vec{P}_{NLOS}$ is the power profile of the NLOS components.
 Finally, additive white gaussian noise, determined by a given SNR value, is added to $\vec{P}_{TOT}$ to create the final augmented training data. This augmentation process, shown in Figure~\ref{fig:augmentation}, is repeated at every training iteration to create new, diverse data for the BeamSeek MLP.
\section{Experimental Evaluation}

\subsection{Hardware Platform}

The experiment utilizes the Sandbox 1 (SB1) environment in the COSMOS (Cloud Enhanced Open Software Defined Mobile Wireless Testbed for City-Scale Deployment) platform supported by the National Science Foundation's (NSF) Platforms for Advanced Wireless Research (PAWR) program \cite{Chen2021}. This testbed features a pair of Sivers IMA 57-66 GHz WiGig transceivers coupled with Xilinx UltraScale+ ZCU111 RFSoC devices that provide baseband signal processing capabilities, allowing real-time experimentation with programmable mmWave systems. \par Most importantly, the Sivers IMA phased arrays at SB1 are mounted on movable XY tables, which allows the relative position and the physical angle ($\alpha$) of the RX array to be adjusted. Figure~\ref{fig:table} includes a close-up view of the Sivers IMA WiGig transceiver \cite{Chen2021}, demonstrating the hardware constraint. It only provides the combined In-phase/Quadrature (I/Q) data across the entire array, as opposed to the I/Q data of each element in the array. Figure~\ref{fig:table} also shows the array mounted on a moving XY table and the surrounding environment. In contrast to the evaluation of previous methods, which were only tested in simulation or RF anechoic chambers, the conditions of this testbed better reflect real-world deployments. Although there is an unobstructed LOS between the arrays, there are multiple structures and radios which may generate interference and multipath effects. \par

\subsection{Data Acquisition}
In the experiment, the arrays are first moved to face each other head-on. The transmit array remains stationary while continuously sending QAM-modulated random data. During the transmission, the receive array is rotated over the full azimuthal range ($-45^\circ$ to $45^\circ$). At each $\alpha$, an exhaustive sweep of all 64 beams in the Sivers IMA transceiver's integrated beambook is performed, receiving 512 samples of raw I/Q data per beam. This set of raw I/Q data, obtained at a very high SNR level (40 dB), forms the basis of the training set. The dataset is repeatedly sliced based on the number of scan beams, returning equally spaced beam indices across the total scan range, as well as segments of the raw I/Q samples for each beam. The gain profiles of each DOA can be computed based on the variance of the I/Q data, which is later augmented during training. \par
The test dataset is obtained in a similar fashion, but the beam index is rapidly switched to receive a total of 512 samples across the desired number of beams $B$ to emulate a deployed version of this method. Twenty test gain profiles were collected from the testbed at each of the SNR ranges outlined in Table I. The DOA estimation error, defined as $\epsilon = \alpha-\hat{\alpha}$, is averaged to produce the results in Section V.

\begin{table}[h]
    \centering
    \caption{SNR Values and Corresponding Labels}
    \label{tab:snr_labels}
    \begin{tabular}{|c|c|}
        \hline
        \textbf{Peak SNR (dB)} & \textbf{Label} \\
        \hline
        0   & Low SNR \\
        5   & Mid-Low SNR \\
         10  & Mid SNR \\
        15  & Mid-High SNR \\
        \hline
    \end{tabular}
\end{table}

\section{Results and Discussion}

\begin{figure}
    \centering
    \includegraphics[width=\linewidth]{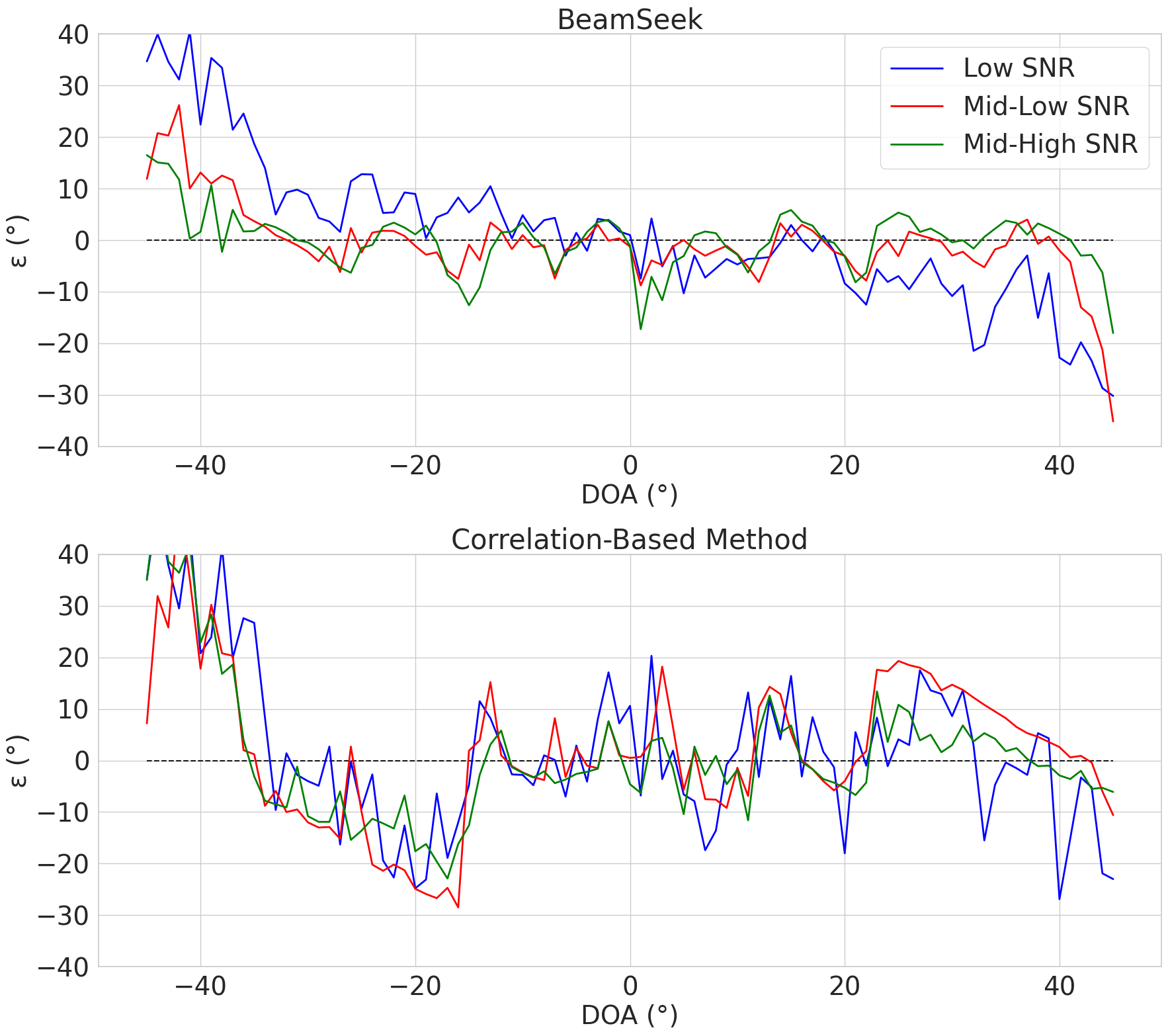}
    \caption{Average DOA estimation error in degrees for the correlation-based method and BeamSeek with eight scan beams ($B=8$)}
    \label{fig:corrmlp}
\end{figure}

\begin{figure}
    \centering
    \includegraphics[width=\linewidth]{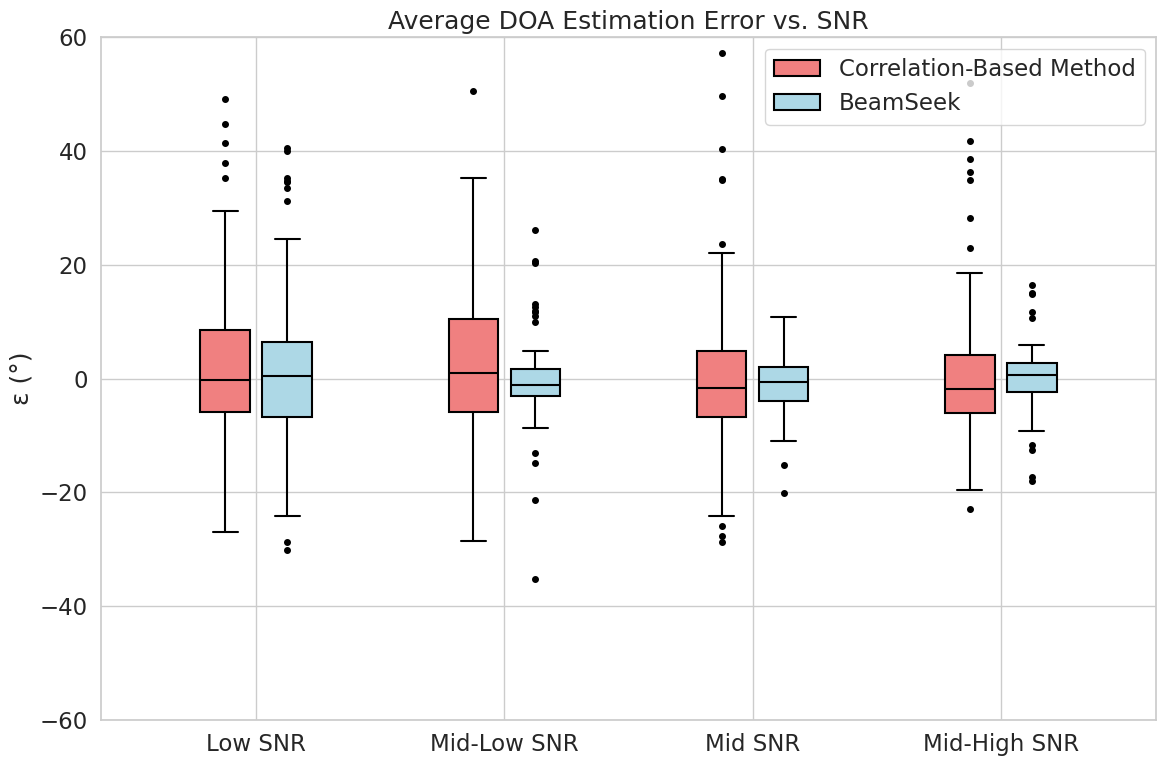}
    \caption{Average DOA estimation error in degrees at every SNR range for correlation-based method and BeamSeek with $B=8$.}
    \label{fig:eVsSnr}
\end{figure}
\begin{figure}
    \centering
    \includegraphics[width=\linewidth]{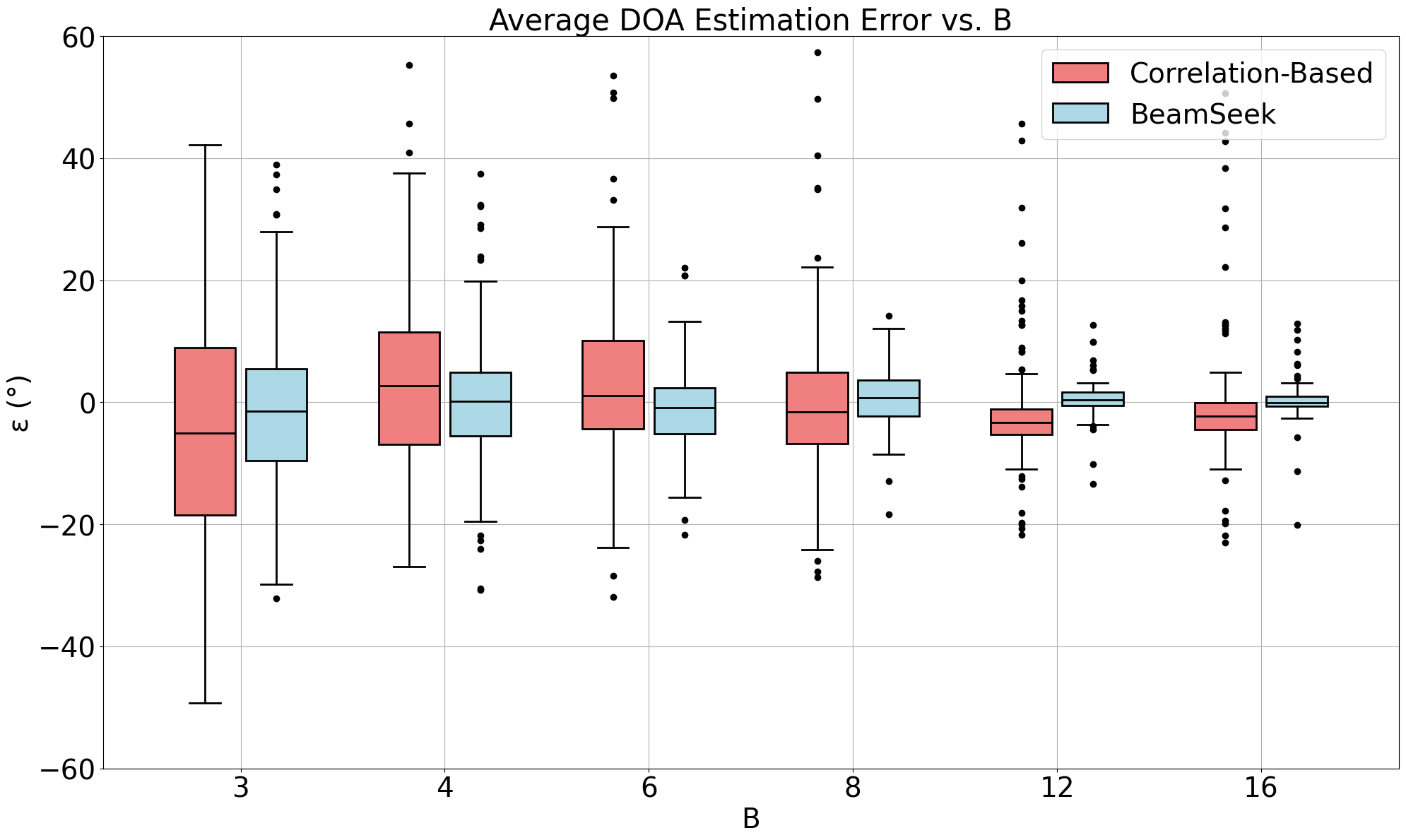}
    \caption{Average DOA estimation error in degrees with a varying number of beams $B$ for correlation-based method and BeamSeek with Mid-SNR.}
    \label{fig:eVsB}
\end{figure}

The average error in DOA estimation using the baseline correlation method and BeamSeek for a varying number of beams is summarized in Table II. Figure~\ref{fig:corrmlp} depicts the average angle estimation error for both methods using 8 scan beams. BeamSeek shows a higher accuracy than the agile correlation method, which is expected given that it was calibrated and has a higher computational complexity. The iterative training method was effective, because BeamSeek demonstrated stronger noise resilience, as shown in Figure~\ref{fig:eVsSnr}. BeamSeek also scales down to a smaller number beams better than the correlation method; as shown in Figure~\ref{fig:eVsB}, the difference in the estimation error between three scan beams and sixteen scan beams is smaller. It is important to note that the Sivers IMA transceiver's integrated beambook contains only 64 beams to cover a $91^\circ$ range, so errors as small as $2^\circ$ would have a negligible effect on the performance of the system, especially considering the spread of the beam over distance. BeamSeek's accuracy increases for a greater number of scan beams, but the estimation error for the DOA still increases as the divergence from broadside ($|\alpha|$) increases. This trend suggests that the model is biased to estimate the DOA closer to $0^\circ$ because it minimizes the error across all incorrect predictions. Training the model using a different optimization criterion may decrease this bias.\par

\begin{table}[h]
    \centering
    \caption{Absolute value of average error $|\bar{\epsilon}|$ for number of scan beams $B$ with the correlation method (Corr) and BeamSeek (MLP)}
    \label{tab:my_label}
    \begin{tabular}{|c|cc|cc|cc|}
    \hline
         & \multicolumn{2}{c|}{Low SNR} & \multicolumn{2}{c|}{Mid-Low SNR} & \multicolumn{2}{c|}{Mid-High SNR} \\

         $B$ & \textbf{Corr} & \textbf{MLP} & \textbf{Corr} & \textbf{MLP} & \textbf{Corr} & \textbf{MLP} \\
        \hline
        3  & 21.45$^\circ$ & 14.40$^\circ$ & 19.55$^\circ$ & 11.54$^\circ$ & 16.06$^\circ$ & 9.54$^\circ$ \\
        4  & 21.15$^\circ$ & 12.92$^\circ$ & 18.73$^\circ$ & 10.55$^\circ$ & 17.68$^\circ$ & 11.04$^\circ$ \\
        6  & 14.84$^\circ$ & 9.99$^\circ$  & 12.00$^\circ$ & 6.97$^\circ$  & 10.83$^\circ$ & 9.24$^\circ$ \\
        8  & 11.18$^\circ$ & 7.73$^\circ$  & 7.66$^\circ$  & 4.30$^\circ$  & 5.96$^\circ$  & 3.86$^\circ$ \\
        12 & 8.99$^\circ$  & 6.10$^\circ$  & 5.13$^\circ$  & 3.02$^\circ$  & 4.23$^\circ$  & 2.45$^\circ$ \\
        16 & 7.26$^\circ$ & 5.47$^\circ$  & 4.28$^\circ$  & 2.50$^\circ$  & 4.14$^\circ$  & 1.96$^\circ$ \\
        \hline
    \end{tabular}
\end{table}

\section{Conclusion and Future Work}
In this paper, we introduced BeamSeek, an MLP-based approach for swift DOA estimation that combines agile beam-switching with DL. This approach addresses the limitations of high-complexity hardware techniques that require multiple RF chains, exhaustive beam-sweeping methods with high latency, and correlation-based methods that struggle in noisy environments.
Our experimental results demonstrate that BeamSeek achieves a superior performance to the baseline correlation-based method across various SNRs, achieving up to an $8^\circ$ reduction in average angle estimation error. These advantages make BeamSeek particularly suitable for practical mmWave deployments in noisy environments and hardware-constrained systems, such as drones and vehicles. BeamSeek does not depend on a known pilot, increasing its scope for Signals Intelligence (SIGINT) applications, as well as decreasing its overall latency in systems where the pilot is only transmitted at the start of a frame or symbol. \par
This work can be further expanded in many ways. Implementing a blended approach where a neural network is used to tune the output of the correlation-based method may decrease the computational complexity. Integrating BeamSeek to perform additional physical layer functions, such as channel estimation and equalization, will further increase its practicality. Lastly, adapting BeamSeek for hybrid MIMO systems composed of several analog subarrays will enable it to be implemented at larger scales.
\par
BeamSeek is a new path toward practical, robust DOA estimation for next-generation wireless systems with hardware constraints, contributing to the development of more efficient and reliable mmWave communication links.

\section*{Acknowledgment}
The authors would like to thank the COSMOS team at Rutgers University and Columbia University for providing access to their testbed resources. Specifically, we are grateful to Ivan Seskar for maintaining the mmWave devices used in this experiment. We would also like to thank Dr. Alexandra Gallyas-Sanhueza and her team at IBM T.J. Watson Research Center for pioneering the agile beamforming method. Finally, we would like to acknowledge the National Science Foundation's Platforms for Advanced Wireless Research program for supporting the COSMOS testbed.

\bibliographystyle{IEEEtran}
\bibliography{citations}

\end{document}